\documentstyle[12pt,psfig,epsf]{article}
\newcommand{\be}{\begin{equation}}
\newcommand{\ee}{\end{equation}}

\newcommand{\beqa}{\begin{eqnarray}}
\newcommand{\eeqa}{\end{eqnarray}}

\newcommand{\eqref}[1]{(\ref{#1})}

\def\boxit#1{\vbox{\hrule\hbox{\vrule\kern8pt
\vbox{\hbox{\kern8pt}\hbox{\vbox{#1}}\hbox{\kern8pt}}
\kern8pt\vrule}\hrule}}
\def\mathboxit#1{\vbox{\hrule\hbox{\vrule\kern8pt\vbox{\kern8pt
\hbox{$\displaystyle #1$}\kern8pt}\kern8pt\vrule}\hrule}}

\def\IB{\relax\hbox{$\inbar\kern-.3em{\rm B}$}}
\def\IC{\relax\hbox{$\inbar\kern-.3em{\rm C}$}}
\def\ID{\relax\hbox{$\inbar\kern-.3em{\rm D}$}}
\def\IE{\relax\hbox{$\inbar\kern-.3em{\rm E}$}}
\def\IF{\relax\hbox{$\inbar\kern-.3em{\rm F}$}}
\def\IG{\relax\hbox{$\inbar\kern-.3em{\rm G}$}}
\def\IGa{\relax\hbox{${\rm I}\kern-.18em\Gamma$}}
\def\IH{\relax{\rm I\kern-.18em H}}
\def\IK{\relax{\rm I\kern-.18em K}}
\def\IL{\relax{\rm I\kern-.18em L}}
\def\IP{\relax{\rm I\kern-.18em P}}
\def\IR{\relax{\rm I\kern-.18em R}}
\def\IZ{\relax\ifmmode\mathchoice
{\hbox{\cmss Z\kern-.4em Z}}{\hbox{\cmss Z\kern-.4em Z}}
{\lower.9pt\hbox{\cmsss Z\kern-.4em Z}} {\lower1.2pt\hbox{\cmsss
Z\kern-.4em Z}}\else{\cmss Z\kern-.4em Z}\fi}

\def\II{\relax{\rm I\kern-.18em I}}

\begin{document}

\hfill  NRCPS-HE-03-10

\vspace{5cm}
%\begin{titlepage}
%\title{
\begin{center}
{\LARGE Gauge Fields-Strings Duality\\
and \\
Tensionless Superstrings

}%title ends

\vspace{2cm}
%\author{

{\sl G.K.Savvidy\footnote{savvidy@inp.demokritos.gr\\
}\\
National Research Center Demokritos,\\
Ag. Paraskevi, GR-15310 Athens, Hellenic Republic\\

}%author ends
%}
%\date{}%in order NOT to write the date
%\maketitle
\end{center}
\vspace{60pt}

\centerline{{\bf Abstract}}

\vspace{12pt}

\noindent
The duality map between gauge theories and strings suggests that
when the gauge theory is in the weak coupling regime the dual string
tension effectively tends to zero, $\alpha' \rightarrow \infty$.
This observation of Sundborg and Witten initiates a fresh
interest to the old problem of tensionless limit of standard
string theory and to the description of its genuine symmetries.
We approach this problem formulating tensionless string theory
by means of geometrical concept of surface perimeter.
The perimeter action uniquely leads to a tensionless theory.

%\end{abstract}
%\thispagestyle{empty}
%\end{titlepage}

\newpage

\pagestyle{plain}
%\pagenumbering{roman}

\section{Introduction}

It is a longstanding problem to describe low energy behavior
of QCD in terms of string-like extended objects.
Naive identification of standard string theory spectrum with hadronic
spectrum encounters a number of
conceptual problems connected with the appearance
of massless states containing graviton in the string spectrum, soft behavior
of high energy scattering amplitudes, opposite to what one should
expect in parton-like picture of asymptotically free gauge theories
and, not the least, higher dimensional space-time.\\

Essential progress was achieved in
\cite{Maldacena:1997re,Gubser:1998bc,Witten:1998qj,Aharony:1999ti}
where the AdS/CFT correspondence was proposed relating the
classical supergravity approximation $R^{2}_{AdS}/\alpha' \gg 1$
of closed IIB  strings moving on ten-dimensional curved space-time
background  $AdS_5 \times S^5$ with large t'Hooft coupling regime of
${\cal N}=4$ supersymmetric Yang-Mills theory
$R^{4}_{AdS}/\alpha'^{2} = \lambda  \equiv  g^{2}_{YM}N$.
In this duality map
one side is weakly coupled, the other is strongly coupled and
there is a natural prescription relating
physical quantities in string and gauge theories,
in particular, identifying  stringy states corresponding
to the leading Regge trajectory with highly excited gauge
theory operators
\cite{Polyakov:2001af, Berenstein:2002jq,Gubser:2002tv}.
The conjecture was applied to calculate Wilson loops,
anomalous dimensions, etc. on the gauge theory
side at strong coupling regime.\\

If one adopts the strong form of the Maldacena conjecture
stating that two theories are exactly the same for all
values of coupling constants\cite{Maldacena:1997re},
then it is important to understand what string theory is like
in the opposite limit
when the gauge theory is in the weak coupling regime
\cite{Haggi-Mani:2000ru,Sundborg:2000wp,Witten:2000,Klebanov:2002ja} .
In that regime the string tension $T = 1/2\pi \alpha'
= \sqrt{\lambda}/R^{2}_{AdS}$
effectively tends to zero and it is natural to assume that
free gauge theory, $\lambda \ll 1$, corresponds to zero tension string
theory, that is to the string theory at extreme energies
\cite{gross,Moore:1993ns,Isberg:1993av}.
The dual description of weakly interacting gauge theory states
on the boundary, in particular, operators
with minimal twist, consisting of bilinear high spin tensors,
on the $AdS_5$ side are supposed to be expressed
in terms of locally interacting
massless gauge fields of arbitrarily high spin
\cite{Ferrara:1998jm,Haggi-Mani:2000ru,Sundborg:2000wp,
Witten:2000,Vasiliev:1999ba,Sezgin:2001zs,Mikhailov:2002bp,Francia:2002aa}.
The gauge theory correlation
functions on the boundary define a high spin
field theory in the bulk with nontrivial interaction vertices
\cite{Bengtsson:1983pg,Bengtsson:1983pd,Savvidy:1998sb}, and
the resulting holographically dual classical gauge field theory
would be the effective description of the desirable string theory
in the bulk,
celebrated symmetric phase of string theory \cite{gross}.\\

This development initiates a fresh interest to the old problem
of tensionless limit of the standard string theory and to the
description of its genuine symmetries
\cite{gross,Haggi-Mani:2000ru,Sundborg:2000wp,
Isberg:1993av,Savvidy:dv,DeVega:1992tm,Clark:2003wk}.
In recent publications we approached this problem
formulating tensionless strings by means of geometrical
concept of surface perimeter, or its length
\cite{gon1,Savvidy:dv}. The perimeter action
uniquely leads to tensionless theory.
It was suggested that nonlinear world-sheet sigma model
which describes tensionless limit is defined by the following action:
\be\label{funaction}
S =m \cdot L=  m \int d^{2}\zeta
\sqrt{h}\sqrt{K^{ia}_{a}K^{ib}_{b}},
\ee
here $m$ has dimension of mass, $h_{ab}$ is the induced metric and
$K^{i}_{ab}$ is the second fundamental form (extrinsic curvature)
\footnote{This action is essentially different
in its geometrical meaning from the action considered in
previous studies \cite{polykov} where it is proportional
to the spherical angle and has dimensionless coupling constant.}.
Instead of being proportional to the area of the surfaces, as it
is the case in the standard string theory:
$$
A  \simeq {1\over 2\pi \alpha '}\int
\sqrt{1 - V^{2}_{\bot}} ~dsdt,
$$
the perimeter action (\ref{funaction}) is proportional
to the length of the surface.
Due to the last property the model
has two desirable features. First of all, when the surface degenerates
into a single world line, the perimeter action (\ref{funaction}) becomes
proportional to the length of the world line \cite{gon1,Savvidy:dv}
\be\label{limit}
S ~~\simeq m \int\sqrt{1 - V^{2}_{\bot}}\cdot K(s,t)  ~dsdt
 \rightarrow ~~ m~ \int  \sqrt{1 - V^{2}_{\bot}}~2\pi ~dt,
\ee
where $K(s,t) = d\varphi /ds$ is a string curvature.
Thanks to this property the functional integral
over surfaces simply transforms
into the Feynman path integral for a point-like relativistic particle,
naturally extending it to relativistic strings. Secondly,
the action is equal to the perimeter of the flat Wilson loop $S=m(R+T)$,
where R is space distance between quarks,
therefore at the {\it classical level string tension is equal to zero}.
The action(\ref{funaction}) can be written in the equivalent
form \cite{gon1,Savvidy:dv}
\begin{equation}\label{conaction}
S= {m\over\pi}\int d^{2}\zeta \sqrt{h}\sqrt{ \left(\Delta(h)
X_{\mu}\right)^{2}},
\end{equation}
where ~$h_{ab}=\partial_{a}X_{\mu}\partial_{b}X_{\mu}$ ~is the induced
metric,~$\Delta(h)= 1/\sqrt{h}~\partial_{a}\sqrt{h}h^{ab}
\partial_{b}$ ~is Laplace operator, \footnote{The equivalence follows
from the relations: $K^{ia}_{a} n^{i}_{\mu}= \Delta(h) X_{\mu}$,
where $n^{\mu}_{i}$  are the normals and
$K^{ia}_{a}K^{ib}_{b}=\left(\Delta(h) X_{\mu}\right)^{2},
\quad i,j=1,2,...,D-2$}
$a,b=1,2; \quad \mu=0,1,2,...,D-1$.\\

In the present work I shall
consider so called model {\it B} \cite{Savvidy:dv},
in which two fields $h_{ab}$- the world-sheet metric and $X^{\mu}$ -
the embedding  field are considered as
independent, that is we abandon  the relation
~$h_{ab}=\partial_{a}X_{\mu}\partial_{b}X_{\mu}$ ~ between them.
At this stage there is no direct relation of the model with
embedding into space-time and the model can be considered as
two-dimensional quantum gravity interacting
with scalar fields $X^{\mu}$. We shall refer to the original theory,
where fields are not independent, as to model {\it A}. The interrelation
between them is considered in \cite{Savvidy:dv}.\\

We shall fix the conformal gauge $h_{ab}=\rho\eta_{ab}$ using the
reparametrization invariance of the action (\ref{conaction})
and represent it in two equivalent forms  \cite{Savvidy:dv}
\be\label{gaga}
S={m\over\pi} \int d^{2}\zeta \sqrt{\left(\partial^{2}
X \right)^{2}} ~~\Leftrightarrow~~\acute{S} =
{1\over\pi}\int d^{2}\zeta \{~\Pi^{\mu}~\partial^{2} X^{\mu}
-  \Omega ~ ( \Pi^{2} -m^2)~\},
\ee
where we have introduced the independent field $\Pi^{\mu}$ and
the Lagrange multiplier  $\Omega$. The system of equations
which follows from $\acute{S}$
\beqa\label{orig}
(I)~~~~~~~~~~~~\partial^2 \Pi^{\mu} =0,~~~~~~~~
\partial^2 X^{\mu} - 2\Omega \Pi^{\mu} =0,~~~~~~~~
\Pi^{\mu}\Pi_{\mu}~ = m^{2}
\eeqa
is equivalent to the original equation for $X^{\mu}$
which follows from $S$
\be\label{confequa1}
(I')~~~~~~~~~~~~~~~~\partial^{2}\left(m \frac{\partial^{2}X_{\mu}}
{\sqrt{\left(\partial^{2}X \right)^2}}\right) =0.
\ee
The
$\Pi^{\mu}$ field has the form
$
\Pi^{\mu} = m  \partial^{2}X^{\mu} /
\sqrt{\left(\partial^{2}X \right)^2}.
$
In addition to the reparametrization invariance the system
exhibits a new gauge symmetry.
For a given parametrization the additional invariance has the form
\cite{Savvidy:dv}
$$
\partial^{2}X^{\mu} \rightarrow \partial^{2}X^{\mu} + 2\omega \Pi^{\mu}
,~~~\Pi^{\mu} \rightarrow \Pi^{\mu},~~~~~\Omega \rightarrow \Omega + \omega
$$
where $\omega = \partial_a \omega^a$ and $\omega^a$ is
arbitrary vector field on the world-sheet.
The above gauge transformation
\be\label{newgauge}
\partial^{2} X^{'}_{\mu} = \left(1 +   \frac{ 2\omega }
{\sqrt{\left(\partial^{2}X \right)^2}}\right) ~\partial^{2} X_{\mu}
\ee
defines a set of  fields $X_{\mu}$ describing the same
physics and can be seen as the gauge orbit of this extra
symmetry. Notice that fields on a given gauge orbit are not
related by reparametrization.  This gauge symmetry
renders the  string space-time coordinate $X^{\mu}$ "less"
physical, not gauge invariant observable. Instead, the string
momentum $P^{\mu}=\partial_0 \Pi^{\mu}$ is a gauge
invariant quantity because $\Pi^{'}_{\mu} = \Pi_{\mu}$.\\

Quantization of the bosonic string {\it B} and its massless spectrum
has been derived in \cite{Savvidy:dv}.
The absence of conformal anomaly requires the space-time to be 13-dimensional
$D_c = 13$. In this string theory
{\it all particles, with arbitrary large integer spin, are massless}. This
pure massless spectrum is consistent with the tensionless character
of the model and it was conjectured in \cite{Savvidy:dv}
that it may describe unbroken phase of standard string theory  when
$ \alpha^{'} \rightarrow \infty$ and all masses tend to zero
$M^{2}_{n} = {1\over \alpha^{'}}(n-1) \rightarrow 0$ \cite{gross} .\\

Supersymmetric extension of the model {\it B}
with ${\cal N}=1$ world-sheet supersymmetry
was constructed in \cite{Nichols:2002ux}.
Here I shall demonstrate that actually
it possesses enhanced fermionic symmetry which elevates
${\cal N}=1 $ world-sheet supersymmetry to  ${\cal N}=2 $ world-sheet
supersymmetry. Indeed quantization of the supersymmetric model
shows that its gauge algebra is identical with the well known
${\cal N}=2 $ world-sheet superalgebra \cite{Ademollo:1974fc}.
This new field theory realization of the ${\cal N}=2$
algebra is free from old problem
\cite{D'Adda:1987rx,Ooguri:1990ww,Ooguri:1991fp,
Bershadsky:1993cx,Bershadsky:1993ta}
connected with the introduction of
second  space-time coordinate  $Y^{\mu}$, which was
introduced in addition to the coordinates
$X^{\mu}$ \cite{Ademollo:1974fc}.
Instead, in this model we have naturally
two left-movers $q^{\mu}_{1}$ and $q^{\mu}_{2}$ and two right-movers
$\tilde{q}^{~\mu}_{1}$ and $\tilde{q}^{~\mu}_{1}$ of the $X^{\mu}$ field
\cite{Savvidy:dv}
\beqa\label{ritghleft}
X^{\mu}_{L} = x^{\mu} +
{1\over m}\pi^{\mu}\zeta^{+} +
\sum^{\infty}_{n=1}\sqrt{{2\over n m^2}}~ \{q^{\mu}_{1n}\sin(n\zeta^{+}) +
q^{\mu}_{2n}\cos(n\zeta^{+})\},\nonumber\\
X^{\mu}_{R} =   x^{\mu}
+  {1\over m}\pi^{\mu}\zeta^{-} +
\sum^{\infty}_{n=1}\sqrt{{2\over  n m^2}}~ \{\tilde{q}^{~\mu}_{1n}\sin(n\zeta^{-}) +
\tilde{q}^{~\mu}_{2n}\cos(n\zeta^{-})\}.\nonumber
\eeqa
The conjugate field is described by a separate field $\Pi^{\mu}$,
\beqa
\Pi^{\mu}_{L}=   m e^{\mu} +
 k^{\mu}\zeta^{+} + \sum^{\infty}_{n=1}\sqrt{{2m^2\over n}}
\{ -p^{\mu}_{1n}\cos(n\zeta^{+}) + p^{\mu}_{2n}\sin(n\zeta^{+})\},\nonumber\\
\Pi^{\mu}_{R}=   m e^{\mu} +
k^{\mu}\zeta^{-} + \sum^{\infty}_{n=1}\sqrt{{2m^2\over n}}
\{-\tilde{p}^{\mu}_{1n}\cos(n\zeta^{-}) + \tilde{p}^{\mu}_{2n}\sin(n\zeta^{-})\},
\nonumber
\eeqa
therefore we have two times more degrees of freedom than in the
standard bosonic string theory. This result can be qualitatively
understood if one takes into account the fact that the field
equations here are of the fourth order (\ref{confequa1}).
Notice that there is also doubling of zero modes,
the new zero mode coordinates are $e^{\mu}$ and their
conjugate variables are $\pi^{\mu}$, they
describe transversal polarizations \cite{Savvidy:dv}.\\

In the first part of this article I shall review the quantization of the
bosonic tensionless string and shall describe its symmetries.
In the second part
I shall present oscillator representation of the supersymmetric
extension of the model and its quantization. In the last section the
twisted topological string model will be constructed in analogy with the
standard prescription for ${\cal N }=2 $ superconformal
field theories
\cite{witten,Labastida:1988zb,Montano:1988dr,eguchi,Ooguri:1995cp}.

\section{Closed Bosonic Strings}

In this section I shall review some facts concerning solution and
quantization of the closed bosonic string which was defined in the previous
section and shall discuss algebraic structure of the
corresponding gauge symmetries (\ref{oldalgebra}),(\ref{newalgebra1})
and (\ref{newalgebra}).
As we shall see they naturally contain Virasoro algebra as its subalgebra
and additional new generators $\Theta_{nk}$ associated with
new gauge symmetry (\ref{newgauge}) forming an Abelian subalgebra.
The conformal algebra has here its
classical form with twice larger central
charge $2 \times {D\over 12} = {D\over 6}$.
This result can be qualitatively understood if one takes
into account the fact that the field equations here are of the fourth order
and therefore we have two left and
two right movers of $X^{\mu}$ field,
two times more degrees of freedom than in the
standard bosonic string theory. Therefore it is not surprising
that the absence of conformal anomaly requires the space-time
to be 13-dimensional: $D_c = 13$.\\

For the closed bosonic strings the mode expansion of $X$ field
(\ref{ritghleft}) can be written in the form \cite{Savvidy:dv}:
\beqa\label{decom}
X^{\mu}_{L} = x^{\mu} +
{1\over m}\pi^{\mu}\zeta^{+} +
i\sum_{n \neq 0}  {1\over n }~ \beta^{\mu}_{n} e^{-in\zeta^{+}},\nonumber\\
X^{\mu}_{R} =   x^{\mu}
+  {1\over m}\pi^{\mu}\zeta^{-} +
i\sum_{n \neq 0}  {1\over n }~ \tilde{\beta}^{\mu}_{n} e^{-in\zeta^{-}},\nonumber
\eeqa
where
$
X^{\mu}= {1\over 2}(X^{\mu}_{L}(\zeta^+) + X^{\mu}_{R}(\zeta^-)),
$
and in similar manner
$\Pi^{\mu}  = {1\over 2}(\Pi^{\mu}_{L}(\zeta^{+})
+ \Pi^{\mu}_{R}(\zeta^{-}))
$
\beqa
\Pi^{\mu}_{L}=   m e^{\mu} +  k^{\mu}\zeta^{+} +
i\sum_{n \neq 0}  {1\over n }~ \alpha^{\mu}_{n} e^{-in\zeta^{+}},\nonumber\\
\Pi^{\mu}_{R}=   m e^{\mu} + k^{\mu}\zeta^{-} +
i\sum_{n \neq 0}  {1\over n }~ \tilde{\alpha}^{\mu}_{n} e^{-in\zeta^{-}}.
\eeqa
The nonzero commutator is
$
[\partial_{\pm} X^{\mu}_{L,R}(\zeta),
\partial_{\pm}\Pi^{\nu}_{L,R}(\zeta^{'})]= 2\pi i~ \eta^{\mu\nu} \delta^{~'}
(\zeta  - \zeta^{'}),
$
with all others equal to zero.
The momentum density operator is $2P^{\mu}  = \partial_{+}\Pi^{\mu}_{L} +
\partial_{-}\Pi^{\mu}_{R} = P^{\mu}_{L} + P^{\mu}_{R}$.
This canonical commutation among the fields imply also
the following commutation relations among the coefficients of
the Fourier expansion (\ref{decom}):
\beqa
[e^{\mu}, \pi^{\nu}]=[x^{\mu}, k^{\nu}] =  i\eta^{\mu\nu},~~~
~[\alpha^{\mu}_{n}, \beta^{\nu}_{k}]= n~\eta^{\mu\nu}\delta_{n+k,0}
\eeqa
and similar ones for $\tilde{\alpha}^{\mu}_{n}$ and
$\tilde{\beta}^{\mu}_{n}$. All other commutators are equal to zero.
 They are connected with the creation and
annihilation operators in the following way
\beqa
\alpha^{\mu}_{n}&=& m \sqrt{n}~a^{\mu}_{n},
~~~ n > 0,~~~~~\alpha^{\mu}_{0}=k^{\mu};~~~
\beta^{\mu}_{n}= {1\over m} \sqrt{n}~b^{\mu}_{n},~~~
n > 0,~~~~~\beta^{\mu}_{0}=\pi^{\mu}/m  \nonumber\\
\alpha^{\mu}_{-n}&=& m \sqrt{n}~a^{+\mu}_{n},
~~~ n > 0,~~~~~~~~~~~~~~~~~~~
\beta^{\mu}_{-n}= {1\over m} \sqrt{n}~b^{+\mu}_{n},~~~
n > 0,
\eeqa
with nonzero commutator
$
[a^{\mu}_{n},b^{+\mu}_{m}]= \eta^{\mu\nu} \delta_{nm}.
$

The Virasoro operators $L_n$  and
new operators $\Theta_{n,k}$ are defined as
\be\label{constcompo}
L_{n}  = <e^{in\zeta^+} :P^{\mu}_{L}~\partial_{+}X^{\mu}_{L}: >,~~~~
\Theta_{n,l}  =  <e^{in\zeta^+ + il\zeta^-} :  \Pi^{\mu}~\Pi^{\mu} -m^2  :>
\ee
and have the form
\beqa\label{oscillators}
L_{n} &=&\sum_{l}:\alpha_{n-l} \cdot\beta_{l}:~~~~~
\tilde{L}_{n} =\sum_{l}: \tilde{\alpha}_{n-l} \cdot \tilde{\beta}_{l}:\nonumber\\
\Theta_{0,0} &=& m^2( e^{2} -1)
+ \sum_{n \neq 0}
{1\over 4 n^2}:(\alpha_{-n}~\alpha_{n} +
\tilde{\alpha}_{-n}~\tilde{\alpha}_{n}):\nonumber\\
\Theta_{n,0} &=& {im\over n}~e \cdot \alpha_{n}
-{1\over 4}\sum_{l \neq 0,n}
{1\over (n-l)l}:\alpha_{n-l}\cdot\alpha_{l}:~~~~~~~~~~~n=\pm1,\pm2,..\nonumber\\
\Theta_{0,n} &=&{im\over n}~e \cdot \tilde{\alpha}_{n}
-{1\over 4}\sum_{l \neq 0,n}
{1\over (n-l)l}:\tilde{\alpha}_{n-l}\cdot\tilde{\alpha}_{l}:
~~~~~~~~~~~n=\pm1,\pm2,..\nonumber\\
\Theta_{n,k} &=& -{1\over 2 n k}:\alpha_{n}\cdot\tilde{\alpha}_{k}:
~~~~~~~~~n,k= \pm 1, \pm 2,....
\eeqa
The conformal algebra has here its classical form but
with twice larger central charge
\be\label{oldalgebra}
[L_n ~,~ L_k] = (n-k) L_{n+k} + {D\over 6}(n^3 -n)\delta_{n+k,0}
\ee
and with the similar expression for right movers $\tilde{L}_n$.
The reason that the central charge is twice bigger than in the
standard bosonic string theory $2 \times {D\over 12} = {D\over 6}$
is simply because we have two left and
two right movers of $X_{\mu}$ field.
Such doubling of modes is reminiscent to the bosonic part of the
${\cal N}=2$ superstring \cite{Ademollo:1974fc}. In the last model there was an
essential problem in identifying the $Y^{\mu}$ coordinates which are
introduced in addition to the normal coordinates $X^{\mu}$
\cite{Ademollo:1974fc}.
In our model the coordinate field $X$ has simply
two sets of commuting oscillators and the conjugate fields are
described by the separate field $\Pi$ .

The full extended gauge symmetry algebra of constraints
(\ref{constcompo}) takes the form
\beqa\label{newalgebra1}
~[L_n  ~, \Theta_{0,0}] &=& -2n  \Theta_{n,0}~~~~~~~~~~~~~~~~~~~~~~~~~~~~
[\tilde{L}_n  ~, \Theta_{0,0}] = -2n \Theta_{0,n} \nonumber\\
~[L_n , \Theta_{k,0}] &=& -(n+k) \Theta_{n+k,0}~~~~~~~~~~~~~~~~~~~
[\tilde{L}_n , \Theta_{k,0}] = -2n \Theta_{k,n}\nonumber\\
~[L_n , \Theta_{0,k}] &=& -2n\Theta_{n,k}~~~~~~~~~~~~~~~~~~~~~~~~~~~~
[\tilde{L}_n ,  \Theta_{0,k}]
= -(n+k) \Theta_{0,n+k} \nonumber\\
~[L_{n} , \Theta_{k,l}] &=& -(n+k) \Theta_{n+k,l}~~~~~~~~~~~~~~~~~~~
[\tilde{L}_n , \Theta_{k,l}]
= -(n+l) \Theta_{k,n+l} \nonumber\\
\eeqa
and one should stress that it is an essentially Abelian extension
\be\label{newalgebra}
 [\Theta_{n,k} , \Theta_{l,p}] =
0,~~~~~n,k,l,p=0, \pm 1,\pm 2,...
\ee
One can easily check that Jacobi identities  between all
these operators are satisfied, therefore the relations
(\ref{oldalgebra}), (\ref{newalgebra1}) and (\ref{newalgebra})
define Abelian extension of Virasoro algebra. The equations
(\ref{oscillators}) suggest its oscillator representation.

To define the physical Hilbert space we should first
impose the Virasoro constraints
\beqa\label{physicalh}
(L_0 +a)\Psi_{phys} &=&0\\
%\{ {k \cdot \pi \over m }  +\sum^{\infty}_{n=1}
%n(a^{+}_n b_n + b^{+}_n a_n) +a\}\Psi_{phys}=0\\
L_n \Psi_{phys} &=&
%\{\sqrt{n}~ {k \cdot b_n \over m} + \sqrt{n} ~\pi \cdot a_n
%+ \sum^{n-1}_{k=1} \sqrt{k(n-k)}  a_{k} \cdot b_{n-k}  \nonumber\\
%&+& \sum^{\infty}_{k=1} \sqrt{k(n+k)}(  a^{+}_{k} \cdot b_{n+k} +
% b^{+}_{k} \cdot a_{n+k}) \}
%\Psi_{phys} =0
0,~~~~~~n =1,2..\nonumber
\eeqa
and then our new constraint $\Theta$. The last operator
has a linear and quadratic $\tau$ dependence which in fact
uniquely define the spectrum of this string theory
$$
(\Pi^2 -m^2) = k^2 ~\tau^2 + 2\{m e \cdot k ~
+ ~ k\cdot\Pi_{oscil} \} \tau + \Pi^{2}_{oscil} +
2m e \cdot \Pi_{oscil}
+ m^2( e^2 -1).
$$
Indeed the first operator diverges quadratically with $\tau$
and the second one linearly. Therefore in order to have
normalizable states in physical Hilbert-Fock space
one should impose corresponding constraints.
We are enforced to define the physical Hilbert space as
\be\label{newconst}
k^2 ~ \Psi_{phys} =0,~~~e \cdot k ~\Psi_{phys} =0,
~~~k \cdot \alpha_{n}  ~\Psi_{phys} =0,~~~k \cdot
\tilde{\alpha}_{n} ~\Psi_{phys} =0,~~~~~n>0.
\ee
The first equation states that {\it all physical states with different spins
are massless}. This is consistent with the tensionless character
of the theory. The rest of the constraints take the form
\beqa\label{newconstraints}
%\Theta_{0,0} \Psi_{phys}&=&0\nonumber\\
%\Theta_{n,0}\Psi_{phys} &=& 0~~~~~~~~~~~~n= 1,2,....\nonumber\\
%\Theta_{0,n}\Psi_{phys} &=&0 ~~~~~~~~~~~~n= 1,2,....\nonumber\\
\Theta_{n,k}\Psi_{phys}  &=&0 ~~~~~~~~~~~~n,k= 0,1,2,....
%{1\over m^2}\Theta_{0,0} \Psi_{phys}  &=&
%\{( e^{2} -1) + \sum^{\infty}_{n=1}
%{1\over 2n}(a^{+}_{n}~a_{n} +
%\tilde{a}^{+}_{n}~\tilde{a}_{n}) \}\Psi_{phys}= 0
%{1 \over m^2}\Theta_{n,0}\Psi_{phys} &=& \{{i\over \sqrt{n}} e \cdot a_{n}
%-{1\over 4} \sum^{n-1}_{l=1}
%{a _{l}~a_{n-l}\over   \sqrt{l(n-l)}}  -{1\over 2} \sum^{\infty}_{l=1}
%{a^{+}_{l}~a_{n+l}\over \sqrt{l(n+l)}}  \}\Psi_{phys} =0\nonumber\\
%{1\over m^2}\Theta_{0,n}\Psi_{phys} &=& \{{i\over \sqrt{n}}e \cdot \tilde{a}_{n}
%-{1\over 4} \sum^{n-1}_{l=1}
%{ \tilde{a} _{l}~\tilde{a}_{n-l} \over \sqrt{l(n-l)} } -
%{1\over 2} \sum^{\infty}_{l=1}
%{\tilde{a}^{+}_{l}~\tilde{a}_{n+l} \over \sqrt{l(n+l)} }\}\Psi_{phys} =0\nonumber\\
%{1\over m^2}\Theta_{n,k}\Psi_{phys}  &=& \{ - {1\over 2\sqrt{nk}} a_{n}~\tilde{a}_{k}
%\}\Psi_{phys} =0 ~~~~~~~~~~~~n,k= 1,2,....
\eeqa
Thus the the physical Hilbert space is defined by the equations
(\ref{physicalh}), (\ref{newconst}) and (\ref{newconstraints}).
In the next section we shall consider ${\cal N}=1$
world-sheet supersymmetric extension of the
above model \cite{Nichols:2002ux} and shall demonstrate that it
actually exhibits the ${\cal N}=2$
world-sheet supersymmetry.

\section{${\cal N} = 1$ World-sheet Supersymmetry}

In the recent article \cite{Nichols:2002ux} the authors constructed
the ${\cal N}=1$ supersymmetric extension of the above model using world-sheet
superfields \cite{ramond,neveu,zumino,green,polchinski}.
Both forms of the action (\ref{gaga}) can be extended to the
supersymmetric case as follows.
For the basic fields~$(X,\Pi,\Omega)$ in (\ref{gaga})
one should introduce the corresponding superfields
\beqa
\hat{X}^{\mu} &=& X^{\mu} + \bar{\vartheta} \Psi^{\mu}+ {1\over 2}
\bar{\vartheta} \vartheta F^{\mu}\nonumber\\
\hat{\Pi}^{\mu} &=& \Pi^{\mu} + \bar{\vartheta} \eta^{\mu}+ {1\over 2}
\bar{\vartheta} \vartheta \Phi^{\mu}\nonumber\\
\hat{\Omega}  &=&  \omega  + \bar{\vartheta} \xi + {1\over 2}
\bar{\vartheta} \vartheta \Omega ,
\eeqa
where $\vartheta$ is an anti-commuting variable
and shall define the supersymmetric action
simply exchanging basic fields ~$(X,\Pi,\Omega)$
in (\ref{gaga}) by corresponding superfields
as follows
\be\label{firstordsusy}
S = {-i \over 2\pi} \int d^{2}\zeta d^{2} \theta \{ ~\hat{\Pi}^{\mu}
\bar{D} D \hat{X}^{\mu} - 2\hat{\Omega} ( \hat{\Pi}^{2} -m^2)~\},
\ee
where
\be
D_A = { \partial \over \partial \bar{\vartheta}^A} -
i (\rho^{a} \vartheta)_A \partial_{a},~~~~
\Psi^{\mu}_{A}(\zeta) \equiv \left( \begin{array}{c}
     \Psi^{\mu}_{-}(\zeta)\\
     \Psi^{\mu}_{+}(\zeta)
     \end{array} \right),~\eta^{\mu}_{A}(\zeta)
     \equiv \left( \begin{array}{c}
     \eta^{\mu}_{-}(\zeta)\\
     \eta^{\mu}_{+}(\zeta)
     \end{array} \right),~\xi_{A}(\zeta)
     \equiv \left( \begin{array}{c}
     \xi_{-}(\zeta)\\
     \xi_{+}(\zeta)
     \end{array} \right),
\ee
$\mu$ is a space-time vector index, $A=1,2$ is a
two-dimensional spinor index.
 $\bar{\Psi}^{\mu} = \Psi^{+\mu} \rho^{0}= \Psi^{T\mu} \rho^{0}$ and
$\rho^{\alpha}$ are two-dimensional Dirac matrices
\be
\{ \rho^{a},\rho^{b} \} =-2 \eta^{a b}.
\ee
In Majorana basis the $\rho's$ are given by
\be
\rho^{0} = \left( \begin{array}{cc}
     0&-i\\
     i&0
     \end{array} \right),~~~~\rho^{1} = \left( \begin{array}{cc}
     0&i\\
     i&0
     \end{array} \right)
\ee
and $i\rho^{\alpha}\partial_{\alpha}$ is a real operator.
The two-dimensional chiral fields are defined as
$
\rho^{3}\Psi^{\mu}_{\pm} = \mp \Psi^{\mu}_{\pm},
$
where $\rho^{3}=\rho^{0}\rho^{1}$.
Substituting superfields one can get the following expression
for the action \cite{Nichols:2002ux}
\beqa\label{fullaction}
S =
{1\over \pi} \int d^{2}\zeta ~\{ ~\Pi^{\mu}\partial^2 X^{\mu}
+ i\bar{\eta^{\mu}}\rho^{a} \partial_{a}\Psi^{\mu}
-F^{\mu}\Phi^{\mu}~~\nonumber\\
~~-\Omega (\Pi^2 -m^2)
-  \omega (2\Pi^{\mu}\Phi^{\mu} +
\bar{\eta^{\mu}}\eta^{\mu}) - 2 \Pi^{\mu}~\bar{\eta^{\mu}}\xi~\}.
\eeqa
The equations of motion are:
\beqa\label{fullequationa}
(I)~~~~~~~~~~~~~~~~~~~\Phi^{\mu} =0\nonumber\\
\partial^2 \Pi^{\mu} =0\nonumber\\
i \rho^{a} \partial_{a}\eta^{\mu}=0\nonumber\\
2 \omega \Pi^{\mu} + F^{\mu} =0\nonumber\\
\partial^2 X^{\mu} - 2\Omega \Pi^{\mu} -2\omega \Phi^{\mu}
-2\bar{\eta^{\mu}}\xi=0\nonumber\\
i\rho^{a} \partial_{a}\Psi^{\mu} -2\Pi^{\mu}\xi -2 \omega\eta^{\mu}=0,
\eeqa
and the variation over Lagrange multipliers gives constraints
\beqa\label{fullconstrants}
(II)~~~~~~~~~~~~~~~~~~\Pi^2 -m^2 =0\nonumber\\
2\Pi^{\mu}\Phi^{\mu} +
\bar{\eta^{\mu}}\eta^{\mu}=0\nonumber\\
2 \Pi^{\mu}~\eta^{\mu} =0.
\eeqa
The first equation represents the constrain which
appears in bosonic tensionless string theory
and the last equation represents its fermionic partners.
As we shall see the first one is the analog of
Klein-Gordon equation and the second one is the analog
of the Dirac equation. In some sense they are more important
relations than the motion equations (I).
The SUSY transformation is:
\beqa
\begin{array}{lll}\label{susy}
\delta X^{\mu} = \bar{\epsilon}\Psi^{\mu},\\
\delta \Psi^{\mu} = -i \rho^{a} \partial_{a}
X^{\mu} ~\epsilon  ~+~F^{\mu}~\epsilon ,\\
\delta F^{\mu} = -i \bar{\epsilon} \rho^{a} \partial_{a}\Psi^{\mu},
\end{array}
\begin{array}{lll}
\delta \Pi^{\mu} = \bar{\epsilon}\eta^{\mu},\\
\delta \eta^{\mu} = -i \rho^{a} \partial_{a}
\Pi^{\mu} ~\epsilon  ~+~\Phi^{\mu}~\epsilon ,\\
\delta \Phi^{\mu} = -i \bar{\epsilon} \rho^{a} \partial_{a}\eta^{\mu},\\
\end{array}
\begin{array}{lll}
\delta \omega = \bar{\epsilon}\xi,\\
\delta \xi^{\mu} = -i \rho^{a} \partial_{a}
\omega ~\epsilon  ~+~\Omega~\epsilon ,\\
\delta \Omega = -i \bar{\epsilon} \rho^{a} \partial_{a}\xi,
\end{array}
\eeqa
where the anti-commuting parameter $\epsilon$ is a
two-dimensional spinor \begin{eqnarray}\epsilon \equiv \left(
\begin{array}{c}
     \epsilon_{-}\\
     \epsilon_{+}
     \end{array} \right).\nonumber
\end{eqnarray}
The action (\ref{fullaction}), equations (\ref{fullequationa})
and the constraints (\ref{fullconstrants}) completely define the
system which exhibits the supersymmetry (\ref{susy}).

As we shall see bellow the action (\ref{fullaction})
possesses enhanced fermionic symmetry which elevates
${\cal N}=1 $ world-sheet supersymmetry up to  ${\cal N}=2 $ world-sheet
supersymmetry. It is convenient to work in
light-cone coordinates. In the light-cone coordinates
the action takes the form
\beqa\label{lagran}
S = {2\over \pi} \int d^{2}\zeta ~\{ - 2\Pi^{\mu}
\partial_{+}\partial_{-} X^{\mu}
+ i\eta^{\mu}_{+}\partial_{-}\psi^{\mu}_{+}
+ i\eta^{\mu}_{-}\partial_{+}\psi^{\mu}_{-}
- {1\over 2}F^{\mu}\Phi^{\mu}~~~\nonumber\\
~~-{1\over 2}\Omega (\Pi^2 -m^2)
-  \omega (\Pi^{\mu}\Phi^{\mu} +
 i\eta^{\mu}_{+}\eta^{\mu}_{-})
-i \Pi^{\mu}~\eta^{\mu}_{+}\xi_{-}
+ i \Pi^{\mu}~\eta^{\mu}_{-}\xi_{+}~\}.
\eeqa
and equations of motion can be solved.
As one can see the SUSY solution of equations (\ref{fullequationa}) is:
$$
i) \Omega =\omega  =\xi =0
$$
and the rest of the equations (I) reduce to the following form
\beqa
(I)~~~~~\partial^2 \Pi^{\mu} =0,~~~i \rho^{a} \partial_{a}\eta^{\mu}=0,~~~
\partial^2 X^{\mu} =0,~~~
i\rho^{a} \partial_{a}\Psi^{\mu} =0,~~~F^{\mu}=\Phi^{\mu}=0
\eeqa
and should be accompanied by the constraints
\beqa
(II)~~~~~\Pi^2 -m^2 =0,~~~~\bar{\eta^{\mu}}\eta^{\mu}=0,~~~~
\Pi^{\mu}~\eta^{\mu} =0.
\eeqa
In the light-cone coordinates these  equations  are easy to solve
since they take the from
\beqa\label{allequations}
\partial_{+}\partial_{-} \Pi^{\mu} =0,~~~\partial_{\pm}\eta^{\mu}_{\mp}=0,~~~
\partial_{+}\partial_{-}  X^{\mu} =0,~~~
\partial_{\pm}\psi^{\mu}_{\mp} =0,\nonumber\\
\Pi^2 -m^2 =0,~~~~\eta^{\mu}_{+}\eta^{\mu}_{-} -
\eta^{\mu}_{-}\eta^{\mu}_{+} =0~~~~
\Pi^{\mu}~\eta^{\mu}_{\pm} =0.
\eeqa
The mode expansion of $X$ field
with the appropriate boundary conditions for closed strings
was given above (\ref{decom}).
The solution of fermionic fields can be represented in
the form of mode expansion as well
\beqa
\eta^{\mu}_{+}= \sum c^{\mu}_{n} e^{-i n\zeta^{+}},~~~~~
\psi^{\mu}_{+}= \sum d^{\mu}_{n} e^{-i n\zeta^{+}}\nonumber\\
\eta^{\mu}_{-}= \sum \tilde{c}^{\mu}_{n} e^{-i n\zeta^{-}},~~~~~
\psi^{\mu}_{-}= \sum \tilde{d}^{\mu}_{n} e^{-i n\zeta^{-}}
\eeqa
with the basic anti-commutators
\beqa\begin{array}{lll}
\{ \eta^{\mu}_{\pm}(\zeta),\psi^{\nu}_{\pm}(\zeta^{'})\}= 2\pi
\eta^{\mu\nu} \delta(\zeta - \zeta^{'}),
\end{array}
\eeqa
and all others equal to zero:
$\{ \eta^{\mu}_{\pm}(\zeta),\eta^{\nu}_{\pm}(\zeta^{'})\}= 0,~
\{ \psi^{\mu}_{\pm}(\zeta),\psi^{\nu}_{\pm}(\zeta^{'})\}= 0$.
Substituting the mode expansion into the anti-commutators
requires following relations between modes
\beqa\label{anticomm}
\{c^{\mu}_{n}, d^{\nu}_{k}\}= \eta^{\mu\nu}\delta_{n+k,0},~~~~
\{c^{\mu}_{n}, c^{\nu}_{k}\}=
0,~~~~~\{d^{\mu}_{n}, d^{\nu}_{k}\}= 0,
\eeqa
and similar ones for $\tilde{c}^{\mu}_{n}$ and
$\tilde{d}^{\mu}_{n}$.

\section{Enhanced ${\cal N} = 2$ World-sheet supersymmetry}
Let us now consider conserved currents: energy
momentum tensor and supercurrent
\beqa
T_{ab} &=& ~\partial_{\{a } \Pi^{\mu}~\partial_{b\}} X^{\mu}
+ i \bar{\eta^{\mu}}\rho_{\{a} \partial_{b\}}\Psi^{\mu}  -trace \nonumber\\
J_{a} &=& {1\over 2 }\rho^{b}\rho_{a}
\Psi^{\mu}\partial_{b}\Pi^{\mu}+{1\over 2 }\rho^{b}\rho_{a}
\eta^{\mu}\partial_{b}X^{\mu}
\eeqa
or in the light-cone coordinates
\beqa
T_{++}&=&2\partial_{+}\Pi^{\mu} \partial_{+}X^{\mu} +
i \eta^{\mu}_{+} \partial_{+}\Psi^{\mu}_{+} - {i\over 2}\partial_{+}
( \eta^{\mu}_{+}\Psi^{\mu}_{+}) ,\nonumber\\
T_{--}&=&2\partial_{-}\Pi^{\mu} \partial_{-}X^{\mu} +
i \eta^{\mu}_{-} \partial_{-}\Psi^{\mu}_{-} - {i\over 2}\partial_{-}
( \eta^{\mu}_{-}\Psi^{\mu}_{-}) ,  \nonumber\\
 J_+ &=& 2\partial_{+}\Pi^{\mu}~\Psi^{\mu}_{+} +
2\eta^{\mu}_{+}\partial_{+}X^{\mu},\nonumber\\
J_- &=& 2\partial_{-}\Pi^{\mu}~\Psi^{\mu}_{-} +
2\eta^{\mu}_{-}\partial_{-}X^{\mu},
\eeqa
Substituting solution (\ref{decom}) into the last formulas one can get
\beqa
T_{++}&=&{1 \over 2}~\partial_{+}\Pi^{\mu}_{L}~\partial_{+}X^{\mu}_{L}+
{i\over 2} \eta^{\mu}_{+} \partial_{+}\Psi^{\mu}_{+} - {i\over 2}
\partial_{+}\eta^{\mu}_{+}~\Psi^{\mu}_{+},\nonumber\\
T_{--}&=&{1 \over 2}~\partial_{-}\Pi^{\mu}_{R}~\partial_{-}X^{\mu}_{R}
+ {i\over 2} \eta^{\mu}_{-} \partial_{-}\Psi^{\mu}_{-}
- {i\over 2}\partial_{-}\eta^{\mu}_{-}~\Psi^{\mu}_{-} ,\nonumber\\
J_+ &=& \partial_{+}\Pi^{\mu}_{L}\Psi^{\mu}_{+}  +
\eta^{\mu}_{+} \partial_{+}X^{\mu}_{L},~~~~~~~~~~~~~~~\nonumber\\
J_- &=& \partial_{-}\Pi^{\mu}_{R}\Psi^{\mu}_{-} +
\eta^{\mu}_{-}\partial_{-}X^{\mu}_{R}.
\eeqa
The mode expansion of these currents is equal to
\beqa
L_{n}  &=& <e^{in\zeta^+} T_{++} >= \sum_{l}:\alpha_{n-l} \cdot\beta_{l}:
+\sum_{l}:(l-{n\over 2}) c_{n-l} \cdot d_{l}: \nonumber\\
F_{n}  &=&  <e^{in\zeta^+} J_{+} >=
\sum_{l}\alpha_{n-l} \cdot d_{l}  +
\sum_{l}\beta_{n-l} \cdot c_{l}.
\eeqa
The standard computation of quantum commutation relations
between these currents gives
\beqa
[L_n , L_m ] &=& (n-m) L_{n+m} + {D\over 4} m^3 \delta_{n+m}\nonumber\\
~[L_n , F_m ] &=& ({n\over 2}-m) F_{n+m}\nonumber\\
~\{F_n , F_m \} &=& 2 L_{n+m} + D n^2 \delta_{n+m}.
\eeqa
This is a standard  ${\cal N}=1 $ superalgebra with the central
charge twice larger than in the standard superstring
theory. It is also clear from the expression for
the full supercurrent $J_a$ that its two separate pieces
$G^{1}_a ={1\over 2 }\rho^{b}\rho_{a} \eta^{\mu}\partial_{b}X^{\mu}$
and $G^{2}_a = {1\over 2 }\rho^{b}\rho_{a}\Psi^{\mu}\partial_{b}\Pi^{\mu}$
also represent conserved currents and therefore pointing to the fact
that there should exist
a higher symmetry group. In the light-cone coordinates
these currents have the form
\beqa\label{N2generators}
G^{1}_+ &=& 2\eta^{\mu}_{+}~\partial_{+}X^{\mu},~~~~~~~~~~
G^{2}_+ = 2\partial_{+}\Pi^{\mu}~\Psi^{\mu}_{+},\nonumber\\
G^{1}_- &=& 2\eta^{\mu}_{-}~\partial_{-}X^{\mu},
~~~~~~~~~~G^{2}_- = 2\partial_{-}\Pi^{\mu}~\Psi^{\mu}_{-}.
\eeqa
Mode expansion of these currents is defined as:
\beqa
G^{1}_{n}  &=&  <e^{in\zeta^+} G^{1}_{+} >=
\sum_{l} \beta_{n-l} \cdot c_{l} \nonumber\\
G^{2}_{n}  &=&  <e^{in\zeta^+} G^{2}_{+} >=
\sum_{l} \alpha_{n-l} \cdot d_{l}
\eeqa
therefore
$$
F_n = G^{1}_n + G^{2}_n
$$
These conserved currents form the following algebra
\beqa
~[L_n , G^{1}_m ] &=& ({n\over 2}-m)
G^{1}_{n+m},~~~~~~~~~~~~~~~~~~~~~~~~~~~~~~(J=3/2)\nonumber\\
~[L_n , G^{2}_m ] &=& ({n\over 2}-m)
G^{2}_{n+m},~~~~~~~~~~~~~~~~~~~~~~~~~~~~~~(J=3/2)\nonumber\\
~\{G^{1}_n , G^{1}_m \} &=&  0,~~~~~~~ \nonumber\\
~\{G^{2}_n , G^{2}_m \} &=&  0.
\eeqa
Here J denotes the conformal spin of the corresponding operators.
The anti-commutator $\{ G^{1}_n , G^{2}_m \}$
cannot be computed in closed form
unless we introduce additional current
\beqa\label{u1current}
T_{a} = {1\over 2 }~\bar{\eta}^{\mu} \rho_{a}\Psi^{\mu}
\eeqa
which appears to be also conserved as it is easy to check
using equations of motion. This conserved
current is connected with the $U(1)$
invariance of the action which rotates
fermionic fields. Its components are
$T_+ = -\eta^{\mu}_{+}\Psi^{\mu}_{+},~~T_- =
-\eta^{\mu}_{-}\Psi^{\mu}_{-},~~\partial_{-} T_+ =
\partial_{+} T_- =0,$ and mode expansion is
\beqa
 T_{n}  =  <e^{in\zeta^+} T_{+} > = -\sum_{l}:c_{n-l} \cdot d_{l}:~.
\eeqa
Then we can compute the anti-commutator:
$$
~\{G^{1}_n , G^{2}_m \} = L_{n+m} +{1\over 2}(n-m)T_{n+m} + {D\over 2}n^2 \delta_{n+m,0}
$$
and the rest of the algebra will take the form:
\beqa
~[L_n , T_m ] &=&  -m T_{n+m} ,~~~~~(J=1)\nonumber\\
~[T_n , T_m ] &=& D n \delta_{n+m,0} \nonumber\\
~[T_n , G^{1}_m ] &=& -G^{1}_{n+m}         \nonumber\\
~[T_n , G^{2}_m ] &=& +G^{2}_{n+m}
\eeqa
It is clear now that this is a well known ${\cal N}=2$ superconformal
algebra \cite{Ademollo:1974fc} and that initially implemented
${\cal N}=1$ SUSY transformation has been naturally enhanced
to ${\cal N}=2$ world-sheet supersymmetry. This symmetry can
also be seen if one introduces the ${\cal N}=2$ superfield as
follows
\beqa
\hat{X}^{\mu}(\zeta,\vartheta_1 , \vartheta_2) =
X^{\mu} +  \vartheta_1 \Psi^{\mu}+ \vartheta_2 \eta^{\mu}
 + \vartheta_2 \vartheta_1 \Pi^{\mu}\nonumber\\
=X^{\mu} +  \vartheta_1 \Psi^{\mu} +\vartheta_2 ~(\eta^{\mu}
 +  \vartheta_1 \Pi^{\mu}).
\eeqa
The important point is that there is no natural extensions of the
superfield $\hat{\Omega}$  to ${\cal N}=2$ superfield,
simply because the constrain $\Pi^2 -m^2 =0$ breaks the symmetry
between $X$ and $\Pi$ fields. This can be seen
as a sign that actually the whole system together with constraints
breaks the ${\cal N}=2$ down to ${\cal N}=1$.
This observation makes the computation of critical
dimension more subtle here. In particular it is not obvious at all that
it should be two, as it is the case for ${\cal N}=2$ strings.

Let us now consider the  new superconstraints (II) (\ref{allequations})
which appear in our case :
$\Delta = \Pi^{\mu}~\eta^{\mu}_{\pm}$
\be
\Delta^{\pm}_{n,l} = <e^{in\zeta^+ + il\zeta^-} :\Pi^{\mu}~\eta^{\mu}_{\pm} :>
\ee
or in terms of oscillators
\beqa\label{susyconstraints}
\Delta^{+}_{0,0} &=&  m e\cdot c_0 + \sum_{n \neq 0}
 \alpha_{-n}~c_{n},~~~~~~\Delta^{-}_{0,0} =  m e\cdot \tilde{c}_0
+ \sum_{n \neq 0}  \tilde{\alpha}_{-n}~\tilde{c}_{n},\nonumber\\
\Delta^{+}_{n,0} &=& i\sum_{l \neq 0,n}
\alpha_{n-l}\cdot c_{l},~~~~~~~~~~~~~~
\Delta^{-}_{0,n} =i \sum_{l \neq 0,n}
\tilde{\alpha}_{n-l}\cdot\tilde{c}_{l} , ~~~~n=\pm 1,\pm 2,...\nonumber\\
\Delta^{+}_{n,l} &=&  \alpha_{n}\cdot\tilde{c}_{l},~~~~~~~~~~~~~~~~~~~~~~~~
\Delta^{-}_{n,l} =  \tilde{\alpha}_{n}\cdot c_{l},~~~~n,l=\pm 1,\pm 2,...
\eeqa
The important fact which uniquely defines the spectrum of this
superstring theory is again the $\tau$ dependence of the operators $\Pi^2$
and $\Pi\cdot\eta_{\pm} $
\beqa
&(\Pi^2 -m^2) = k^2 ~\tau^2 + 2\{m e \cdot k ~
+ ~ k\cdot\Pi_{oscil} \} \tau +  m^2( e^2 -1)+
2m e \cdot \Pi_{oscil}+\Pi^{2}_{oscil} ,\nonumber\\
&\Pi\cdot\eta_{+}= (c_0\cdot k  + k\cdot\eta_{+~oscil} ) \tau
+ m e\cdot c_0  + m e \cdot \eta_{+~oscil} +
\Pi_{oscil}\cdot \eta_{+~oscil},\nonumber\\
&\Pi\cdot\eta_{-}= (\tilde{c}_0\cdot k  + k\cdot\eta_{-~oscil} ) \tau
+ m e\cdot \tilde{c}_0  + m e \cdot \eta_{-~oscil} +
\Pi_{oscil}\cdot \eta_{-~oscil}.
\eeqa
The first operator diverges quadratically with $\tau$
and the second one linearly in bosonic sector and we have linear
divergency of first operators in fermionic sector. Therefore in order to have
normalizable states in physical Hilbert space
one should impose corresponding constraints.
We are enforced to define the physical Hilbert space as
\beqa
k^2 ~ \Psi_{phys} =0,~~~c_0\cdot k~\Psi_{phys} =
0,~~~\tilde{c}_0\cdot k~\Psi_{phys} =0,~~~~~~~~~~~~~~~~~~~~~~~\nonumber\\
k \cdot \alpha_{n}  ~\Psi_{phys} =0,~~~k \cdot
\tilde{\alpha}_{n} ~\Psi_{phys} =0,~~~
~~~
k \cdot c_{n}  ~\Psi_{phys} =0,~~~k \cdot
\tilde{c}_{n} ~\Psi_{phys} =0,~~~~~n>0 ,\nonumber\\
e \cdot k ~\Psi_{phys} =0.~~~~~~~~~~~~~~~~~~~~~~~~~~~~~~~~~~~~~~~~~~~~~~~
\eeqa
All these constraints can naturally be grouped  into three systems
of equations. The first three equations
are nothing else but massless  Klein-Gordon and Dirac equations
and uniquely define the spectrum of the theory.
We conclude that {\it all physical states with integer and half
integer spins are massless}. This is consistent with
tensionless character of the theory. The second system of equations
imposes important condition of transversality on fermion
and boson oscillators.  Finally the last equation
suggests that the  vector $e_{\mu}$ should be interpreted
as polarization vector transverse to the momentum vector $k_{\mu}$.

We should  impose  the constraints of ${\cal N}=2$
superconformal algebra
\beqa\label{last}
(L_0 +a )\Psi_{phys} &=&0\nonumber\\
L_n \Psi_{phys}&=&0, ~~~~~~n=1,2,.. \nonumber\\
T_0 \Psi_{phys}&=&0        \nonumber\\
T_n \Psi_{phys}&=&0,~~~~~~n=1,2,.. \nonumber\\
%m L_0 \Psi_{phys} = \{k \cdot \pi  +m \sum^{\infty}_{n=1}
%n(a^{+}_n b_n + b^{+}_n a_n) + m \sum^{\infty}_{n=1}
%n(c^{+}_n d_n + d^{+}_n c_n)  \}\Psi_{phys} &=&0\nonumber\\
%T_0 \Psi_{phys}=\{d_0 c_0 +  \sum^{\infty}_{n=1}
%(d^{+}_n c_n - c^{+}_n d_n)\}\Psi_{phys} &=&0        \nonumber\\
%L_n \Psi_{phys}=0,~~~~~~T_n \Psi_{phys}=0,~~~~~~
G^{1}_n \Psi_{phys}&=&0,~~~~~~n=1,2,.. \nonumber\\
G^{2}_n \Psi_{phys}&=&0,~~~~~~n=1,2,..
\eeqa
together with the additional constraints $\Theta_{k,l}$ (\ref{newconstraints})
and fermionic constraints (\ref{susyconstraints})
\beqa
%\Delta^{\pm}_{0,0}\Psi_{phys}&=&0\nonumber\\
%= \{ c_0 \cdot e  + \sum^{\infty}_{n=1}
%\sqrt{n}(a^{+}_{n}~c_{n} + c^{+}_{n}~a_{n})\}\Psi_{phys}
%\Delta^{\pm}_{n,0}\Psi_{phys}&=&0,~~~~~n=1,2,..\nonumber\\
%\Delta^{\pm}_{0,n}\Psi_{phys}&=&0,~~~~~n=1,2,..\nonumber\\
\Delta^{\pm}_{n,l}\Psi_{phys}&=&0,~~~~~n,l =0,1,2,...
\eeqa
One should study in great details this Hilbert space
in order to learn more about content of the theory
and to prove the absence of the negative norm states.
We cannot also say anything certain about critical
dimension of the model because we  have additional
symmetries and the corresponding constraints, the influence of which
on the calculation of the critical dimension at the moment is not
quite well understood.

In the rest of the article we shall consider a
close topological model which
can be constructed by twisting
\cite{witten,Labastida:1988zb,Montano:1988dr,eguchi}. Indeed the
redefinition of the energy momentum tensor by the total derivative
of the U(1) current, leads to the topological theory.
By this twisting operation the above ${\cal N}=2$
supersymmetry transforms into BRST
symmetry  as in \cite{witten,eguchi}.

\section{Twisted Topological Strings}

We shall obtain the topological version of the above ${\cal N}=2$
theory by the redefinition of the energy momentum tensor T by
$\hat{T}$ as follows:
$$
\hat{T}_{ab} = T_{ab}  - {i\over 2} \partial _{a} J_b,
$$
where $J_b$ is the U(1) current (\ref{u1current}).
In light-cone components we have
\beqa
\tilde{T}_{++}&=&2\partial_{+}\Pi^{\mu} \partial_{+}X^{\mu} +
i \eta^{\mu}_{+} \partial_{+}\Psi^{\mu}_{+}, ~~~~~~~
\tilde{T}_{--}= 2\partial_{-}\Pi^{\mu} \partial_{-}X^{\mu}+
i \eta^{\mu}_{-} \partial_{-}\Psi^{\mu}_{-} , \nonumber\\
F_+ &=& 2\partial_{+}\Pi^{\mu}~\Psi^{\mu}_{+} +
2\eta^{\mu}_{+}\partial_{+}X^{\mu},~~~~~~~~~~
F_-  =2\partial_{-}\Pi^{\mu}~\Psi^{\mu}_{-}+
2\eta^{\mu}_{-}\partial_{-}X^{\mu},\nonumber\\
J_+ &=& - \eta^{\mu}_{+}\Psi^{\mu}_{+},~~~~~~~~~~~~~~~~~~~~~~~~~~~~~~
J_- = - \eta^{\mu}_{-}\Psi^{\mu}_{-},
\eeqa
Substituting solutions (\ref{decom}) into  the last formulas one can get
\beqa
\tilde{T}_{++}&=&{1 \over 2}~\partial_{+}\Pi^{\mu}_{L}~\partial_{+}X^{\mu}_{L}+
i \eta^{\mu}_{+} \partial_{+}\Psi^{\mu}_{+},\nonumber\\
\tilde{T}_{--}&=&{1 \over 2}~\partial_{-}\Pi^{\mu}_{R}~\partial_{-}X^{\mu}_{R}
+ i  \eta^{\mu}_{-} \partial_{-}\Psi^{\mu}_{-},\nonumber\\
F^{1}_+ &=& \partial_{+}\Pi^{\mu}_{L}\Psi^{\mu}_{+}  +
\eta^{\mu}_{+} \partial_{+}X^{\mu}_{L},~~~~~~~~~~~~~~~\\
F^{1}_- &=& \partial_{-}\Pi^{\mu}_{R}\Psi^{\mu}_{-} +
\eta^{\mu}_{-}\partial_{-}X^{\mu}_{R},\nonumber\\
J_+ &=& - \eta^{\mu}_{+}\Psi^{\mu}_{+},  \nonumber\\
J_- &=& - \eta^{\mu}_{-}\Psi^{\mu}_{-},
\eeqa
Mode expansion of these currents is defined as:
\beqa
L_{n}  &=& <e^{in\zeta^+} \tilde{T}_{++} >= \sum_{l}:\alpha_{n-l} \cdot\beta_{l}:
+\sum_{l}: l~  c_{n-l} \cdot d_{l}: \nonumber\\
F^{1}_{n}  &=&  <e^{in\zeta^+} F_{+} >=
\sum_{l}\alpha_{n-l} \cdot d_{l}  +
\sum_{l}\beta_{n-l} \cdot c_{l}  \nonumber\\
F^{2}_{n}  &=&  <e^{in\zeta^+} F^{2}_{+} >=
\sum_{l} \alpha_{n-l} \cdot d_{l}  -
\sum_{l} \beta_{n-l} \cdot c_{l}  \nonumber\\
J_{n}  &=&  <e^{in\zeta^+} J_{+} >=
-\sum_{l}:c_{n-l} \cdot d_{l}:
\eeqa
where we have introduced a new operator $F^{2}_{+}$ which is
equal to the following expression
\be
F^{2}_{+} = \partial_{+}\Pi^{\mu}_{L} \Psi^{\mu}_{+} -
\eta^{\mu}_{+} \partial_{+}X^{\mu}_{L}.
\ee
The necessity of introducing this operator comes from
the fact that again when we compute the algebra between
operators $T,F^{1},J$  the
algebra is closed only if we introduce this new
operator. For these four operators the algebra is closed
\beqa
[L_n , L_m ] &=& (n-m) L_{n+m},\nonumber\\
~[L_n , F^{1}_m ] &=& ({n\over 2}-m) F^{1}_{n+m}
- {n\over 2} F^{2}_{n+m},\nonumber\\
~[L_n , F^{2}_{m}] &=&  ({n\over 2}-m) F^{2}_{n+m}
- {n\over 2}F^{1}_{n+m} ,\nonumber\\
~[L_n , J_m ] &=&  -m J_{n+m} - {D \over 2}n(n+1)\delta_{n+m,0}, \nonumber\\
~[J_n , J_m ] &=& D n \delta_{n+m,0} \nonumber\\
~[J_n , F^{1}_m ] &=& F^{2}_{n+m}         \nonumber\\
~[J_n , F^{2}_{m} ] &=& F^{1}_{n+m}            \nonumber\\
~\{F^{1}_n , F^{1}_m \} &=&2L_{n+m} +(n+m)J_{n+m}
+ D~ n^2\delta_{n+m,0} \nonumber\\
~\{F^{1}_n , F^{2}_{m} \} &=&~~~~~~~~~-(n-m)J_{n+m}
+ D~ n ~\delta_{n+m,0}\nonumber\\
~\{F^{2}_{n}  , F^{2}_{m}  \} &= &
-2L_{n+m} -(n+m)J_{n+m} - D~ n^2\delta_{n+m,0}
\eeqa
There is no defined conformal dimensions for the fermion operators
$F^{1}$ and $F^{2}$, they also do not have defined charges with respect
to the $U(1)$ group. As it is easy to see the linear combination
of these operators do have defined charges. Indeed
we should introduce the linear combination of supercurrents as
we did in  the previous model
\be
2G_n = F^{1}_n - F^{2}_{n},~~~~~2 Q_n =  F^{1}_n + F^{2}_{n},
\ee
in order to have diagonal form of supercurrent with respect to the
conformal operator $L_n$. In coordinate space they look as (\ref{N2generators}).
For these conserved currents
the algebra takes the form
\beqa
[L_n , L_m ] &=& (n-m) L_{n+m},~~~~~~~~~~~~~~~~~~~~~~~~(J=2)\nonumber\\
~[L_n , G_m ] &=& (n-m) G_{n+m},~~~~~~~~~~~~~~~~~~~~~~~~(J=2)\nonumber\\
~[L_n , Q_m ] &=&  -m Q_{n+m},~~~~~~~~~~~~~~~~~~~~~~~~~~~~~~(J=1)\nonumber\\
~[L_n , J_m ] &=&  -m J_{n+m} - {D \over 2}n(n+1)\delta_{n+m,0}~~~~~(J=1)\nonumber\\
~[J_n , J_m ] &=& D m \delta_{n+m,0} \nonumber\\
~[J_n , G_m ] &=& -G_{n+m}         \nonumber\\
~[J_n , Q_m ] &=& +Q_{n+m}            \nonumber\\
~\{G_n , G_m \} &=&  0,~~~~~~~ \nonumber\\
~\{Q_n , Q_m \} &=&  0,~~~~~~~ \nonumber\\
~\{G_n , Q_m \} &= & L_{n+m} +mJ_{n+m} + {D\over 2}n(n+1)\delta_{n+m,0}
\eeqa
which is well known in topological conformal field theory \cite{eguchi}.
One can see that we have here zero central charge and two nilpotent
fermion operators $G$ and $Q$ which form the N=2 world-sheet supersymmetry.
Two conserved supercurrents which appear above should
have come from the explicit fermion symmetry of the theory. As we shall see
in a moment these symmetries can be justified.
One can check that the system is invariant under fermion
transformation laws $\delta$ and $\bar{\delta}$ defined as
follows \cite{Nichols:2002ux}:
\beqa
\begin{array}{llllll}\label{newsusy}
\delta X^{\mu} = 0,\\
\delta \Psi^{\mu}_{-} = -2\epsilon_{+}\partial_{-}X^{\mu} ,\\
\delta \Psi^{\mu}_{+}=0,\\
\delta F^{\mu} =  -2i  \epsilon_{+}   \partial_{-}\Psi^{\mu}_{+},\\
\delta \Pi^{\mu} =  i\epsilon_{+}\eta^{\mu}_{-},\\
\delta \eta^{\mu}_{-} = 0 ,\\
\delta \eta^{\mu}_{+} =  - \epsilon_{+}\Phi^{\mu}~ ,\\
\delta \Phi^{\mu} = 0,
\end{array}~~~~~
\begin{array}{lll}
\bar{\delta} X^{\mu} = 0,\\
\bar{\delta}\Psi^{\mu}_{-} = 0,\\
\bar{\delta }\Psi^{\mu}_{+}= -2\epsilon_{-}\partial_{+}X^{\mu} ,\\
\bar{\delta} F^{\mu} =  2i  \epsilon_{-}   \partial_{+}\Psi^{\mu}_{-},\\
\bar{\delta} \Pi^{\mu} =  i\epsilon_{-}\eta^{\mu}_{+},\\
\bar{\delta}\eta^{\mu}_{-} =  \epsilon_{-}\Phi^{\mu}~ ,\\
\bar{\delta}\eta^{\mu}_{+} = 0 ,\\
\bar{\delta} \Phi^{\mu} = 0,
\end{array}~~~~
\begin{array}{lll}
\delta \omega =  i\epsilon_{+}\xi_{-},\\
\delta \xi_{-} = 0,\\
\delta \xi_{+} =  -\epsilon_{+}\Omega,\\
\delta \Omega = 0,\\
\bar{\delta}\omega = i\epsilon_{-}\xi_{+},\\
\bar{\delta}\xi_{-}= \epsilon_{-}\Omega,\\
\bar{\delta}\xi_{+}=0,\\
\bar{\delta}\Omega= 0,
\end{array}
\eeqa
The  algebra of these fermionic symmetries
is nilpotent and is very similar to BRST transformations
\be\label{brst}
\delta_{\epsilon}\delta_{\acute{\epsilon}} ~ (H)  =
\bar{\delta}_{\epsilon}\bar{\delta}_{\acute{\epsilon}}~ (H)=
0,~~~~~(\delta_{\epsilon} \bar{\delta}_{\acute{\epsilon}} -
\bar{\delta}_{\acute{\epsilon}}\delta_{\epsilon} )~ (H)=0.
\ee
where H is any of the fields $ (X,\Psi,F,\Pi,\eta,\Phi,\omega,\xi,\Omega)$.
From (\ref{brst}) it follows that the action is invariant under
fermionic symmetries (\ref{newsusy}). We can compute  the
current corresponding to this fermion symmetry.
The variation of the action is
\beqa
\delta{S} = {2\over \pi} \int d^{2}\zeta ~\{-2 i\epsilon_{+}\eta^{\mu}_{-}
\partial_{+}\partial_{-}X^{\mu} -i\epsilon_{+}\Phi^{\mu}\partial_{-}\Psi^{\mu}_{+}
+i\eta^{\mu}_{-}\partial_{+}(-2\epsilon_{+}\partial_{-}X^{\mu}) \nonumber\\
-(1/2)(-2i\epsilon_{+}\partial_{-}\Psi^{\mu}_{+})\Phi^{\mu} ~\}=
-{2i\over \pi} \int d^{2}\zeta ~G_{-}~ \partial_{+}(\epsilon_{+})
\eeqa
and supercurrent $G_{-} = 2\eta^{\mu}_{-}\partial_{-}X^{\mu}$
coincides with the one which appeared in the previous section.
The important fact now is that the Lagrangian is
a variation of the super-potentials $W$ and $\bar{W}$
\be
W = \Pi^{\mu}~\partial_{+}\Psi^{\mu}_{-} +
{1\over 2}\eta^{\mu}_{+}~F^{\mu},~~~~~
\bar{W} =  \Pi^{\mu}~\partial_{-}\Psi^{\mu}_{+} -
{1\over 2}\eta^{\mu}_{-}~F^{\mu},
\ee
so that
\be
\delta W = \epsilon_{+}{\cal L},~~~~~~~~~~~~~~~~~~~~~~~~~~~
\bar{\delta} ~\bar{W} =\epsilon_{-}{\cal L}.
\ee
It is also true that there exists a potential $V$
such that
\beqa
\delta ~V= -i\epsilon_{+} \bar{W},~~~~
 ~\bar{\delta} ~V =  i \epsilon_{-} W,~~~~~V= {1\over 2}\Pi^{\mu} F^{\mu},
\eeqa
thus
$$
i\delta \bar{\delta}  V = \epsilon_{+}\epsilon_{-}{\cal L},
~~~~~~~i\bar{\delta} \delta  V = \epsilon_{+}\epsilon_{-}{\cal L}.
$$
The constrains (II) can  also be represented
by the $\bar{\delta} \delta$ transformation and therefore the full
Lagrangian in (\ref{lagran}) can be represented as
\be
2\epsilon_{+}\epsilon_{-}{\cal L}_{tot}
= ~i\bar{\delta} \delta~  \left(  \Pi^{\mu} F^{\mu}
+ \omega (\Pi^2 -m^2)   \right).
\ee
Thus the action can be represented as BRST commutator
${\cal L} = \{ G_+ , W   \}=\{  G_{-},\bar{W }  \}$.
The above fermion symmetry allows to express some important operators
as variation of others. In particular the energy momentum tensor
is a variation of second supercurrent $Q$, but only up to the total
derivative of $U(1)$ current $J_a$
\beqa
\bar{\delta}~Q_{+}&=&-2\epsilon_- T_{++}
+ 2i\epsilon_-\partial_{+}J_{+}  ,\nonumber\\
\delta~Q_{-}&=&-2\epsilon_+ T_{--}  + 2i\epsilon_+\partial_{-}J_{-}  ,
\eeqa
Instead, the supercurrent $G$ introduced above is total variation of the
vector $v_a = \Pi^{\mu}\partial_{a} X^{\mu}$
\beqa
\bar{\delta}~v_+ &=&i\epsilon_- G_+ ,~~~~~~~~~~~~v_+
= 2\Pi^{\mu}\partial_{+} X^{\mu}\nonumber\\
\delta~v_- &=& i\epsilon_+ G_- ,~~~~~~~~~~~~v_-
= 2\Pi^{\mu}\partial_{-} X^{\mu}.
\eeqa
Let us consider the second fermion symmetry of the action
\beqa
\begin{array}{llllll}\label{newsusyII}
\delta X^{\mu} = -i\epsilon_{-}\Psi^{\mu}_{+}~ ,\\
\delta \Psi^{\mu}_{-} = -\epsilon_{-}F^{\mu} ,\\
\delta \Psi^{\mu}_{+}=0,\\
\delta F^{\mu} =  0,\\
\delta \Pi^{\mu} = 0,\\
\delta \eta^{\mu}_{-} = 0 ,\\
\delta \eta^{\mu}_{+} =  2 \epsilon_{-}\partial_{+} \Pi^{\mu}~ ,\\
\delta \Phi^{\mu} = -2i  \epsilon_{-}   \partial_{+}\eta^{\mu}_{-},
\end{array}~~~~~
\begin{array}{lll}
\bar{\delta} X^{\mu} = -i\epsilon_{+}\Psi^{\mu}_{-}~,\\
\bar{\delta}\Psi^{\mu}_{-} = 0,\\
\bar{\delta }\Psi^{\mu}_{+}= \epsilon_{+}F^{\mu} ,\\
\bar{\delta} F^{\mu} =  0,\\
\bar{\delta} \Pi^{\mu} =  0,\\
\bar{\delta}\eta^{\mu}_{-} =  2 \epsilon_{+}\partial_{-} \Pi^{\mu}~ ~ ,\\
\bar{\delta}\eta^{\mu}_{+} = 0 ,\\
\bar{\delta} \Phi^{\mu} = 2i  \epsilon_{+}   \partial_{-}\eta^{\mu}_{+},
\end{array}~~~~
\begin{array}{lll}
\delta \omega =  0,\\
\delta \xi_{-} = -\epsilon_{-}\omega,\\
\delta \xi_{+} = 0 ,\\
\delta \Omega = i\epsilon_{-}\xi_{+},\\
\bar{\delta}\omega = 0,\\
\bar{\delta}\xi_{-}= 0,\\
\bar{\delta}\xi_{+}=\epsilon_{+}\omega,\\
\bar{\delta}\Omega= i\epsilon_{+}\xi_{-},
\end{array}
\eeqa
The  algebra of these fermionic symmetries
is nilpotent and is very similar to BRST transformations
\be\label{brst2}
\delta_{\epsilon}\delta_{\acute{\epsilon}} ~ (H)  =
\bar{\delta}_{\epsilon}\bar{\delta}_{\acute{\epsilon}}~ (H)=
0,~~~~~(\delta_{\epsilon} \bar{\delta}_{\acute{\epsilon}} -
\bar{\delta}_{\acute{\epsilon}}\delta_{\epsilon} )~ (H)=0.
\ee
where H is any of the fields $ (X,\Psi,F,\Pi,\eta,\Phi,\omega,\xi,\Omega)$.
Let us compute  the corresponding
current. The variation of the action is
\beqa
\delta{S} = {2\over \pi} \int d^{2}\zeta ~\{-2
\Pi^{\mu}\partial_{+}\partial_{-}(-i\epsilon_{-} \Psi^{\mu}_{+})
+2i\epsilon_{-}\partial_{+}\Pi^{\mu}\partial_{-}\Psi^{\mu}_{+}
+\nonumber\\
i\eta^{\mu}_{-}\partial_{+}(-\epsilon_{-}F^{\mu}) -(1/2)F^{\mu}
(-2i\epsilon_{-}\partial_{+}\eta^{\mu}_{-})
\}=
-{2i\over \pi} \int d^{2}\zeta ~Q_{+}~ (\partial_{-}\epsilon_{-})
\eeqa
and $Q_{+} = 2\partial_{+}\Pi^{\mu}\Psi^{\mu}_{+}$ also
appeared in the previous section.
The important fact now is that the energy momentum tensor is BRST commutator
with respect to the second fermion symmetry
\beqa
\delta ~G_+   &=& \delta ~(2\eta^{\mu}_{+}\partial_{+}X^{\mu})   =
2\epsilon_{-} T_{++},~~~~\nonumber\\
\delta ~G_-   &=& \delta ~(2\eta^{\mu}_{-}\partial_{-}X^{\mu})   =
2\epsilon_{+} T_{--},
\eeqa
and supercurrent $Q$ ia a variation of $U(1)$ current $J_a$
\beqa
\delta ~J_+   &=& \delta ~(-\eta^{\mu}_{+}\Psi^{\mu}_{+})   =
-\epsilon_{-} Q_{+},\nonumber\\
\delta ~J_-   &=& \delta ~(-\eta^{\mu}_{-}\Psi^{\mu}_{-})   =
-\epsilon_{+} Q_{-}
\eeqa
Thus the energy momentum tensor is a BRST commutator
$T_{++} = \{ Q_+ , G_+  \}$ and its central charge vanishes,
the model transforms into topological theory.

The author wishes to thank A.~D'Adda, I.Bakas,
L.Brink, E.Floratos, M.Vasiliev, R.Manvelyan, A.Nichols and R.Fazio
for stimulating discussions. This work
was supported in part by the EEC Grant no. HPRN-CT-1999-00161 and
NATO Grant PST.CLG.978154.

\vfill

\begin{thebibliography}{99}

\bibitem{Maldacena:1997re}
J.~M.~Maldacena,
%``The large N limit of superconformal field theories and supergravity,''
Adv.\ Theor.\ Math.\ Phys.\  {\bf 2} (1998) 231
[Int.\ J.\ Theor.\ Phys.\  {\bf 38} (1999) 1113]

\bibitem{Gubser:1998bc}
S.~S.~Gubser, I.~R.~Klebanov and A.~M.~Polyakov,
%``Gauge theory correlators from non-critical string theory,''
Phys.\ Lett.\ B {\bf 428} (1998) 105

\bibitem{Witten:1998qj}
E.~Witten,
%``Anti-de Sitter space and holography,''
Adv.\ Theor.\ Math.\ Phys.\  {\bf 2} (1998) 253
[arXiv:hep-th/9802150].

\bibitem{Aharony:1999ti}
O.~Aharony, S.~S.~Gubser, J.~M.~Maldacena, H.~Ooguri and Y.~Oz,
%``Large N field theories, string theory and gravity,''
Phys.\ Rept.\  {\bf 323} (2000) 183

\bibitem{Polyakov:2001af}
A.~M.~Polyakov,
%``Gauge fields and space-time,''
Int.\ J.\ Mod.\ Phys.\ A {\bf 17S1} (2002) 119

\bibitem{Berenstein:2002jq}
D.~Berenstein, J.~M.~Maldacena and H.~Nastase,
%``Strings in flat space and pp waves from N = 4 super Yang Mills,''
JHEP {\bf 0204} (2002) 013

\bibitem{Gubser:2002tv}
S.~S.~Gubser, I.~R.~Klebanov and A.~M.~Polyakov,
%``A semi-classical limit of the gauge/string correspondence,''
Nucl.\ Phys.\ B {\bf 636} (2002) 99

%\cite{Haggi-Mani:2000ru}
\bibitem{Haggi-Mani:2000ru}
P.~Haggi-Mani and B.~Sundborg,
%``Free large N supersymmetric Yang-Mills theory as a string theory,''
JHEP {\bf 0004} (2000) 031

%\cite{Sundborg:2000wp}
\bibitem{Sundborg:2000wp}
B.~Sundborg,
%``Stringy gravity, interacting tensionless strings and massless higher  spins,''
Nucl.\ Phys.\ Proc.\ Suppl.\  {\bf 102} (2001) 113

\bibitem{Witten:2000}
E.~Witten,
Talk at the John Schwarz 60-th Birthday Symposium,
http://theory.caltech.edu/jhs60/witten/1.html

\bibitem{gross} D.Gross, High-Energy symmetries of
string theory, Phys.Rev.Lett. 60 (1988) 1229
%\cite{Moore:1993ns}

\bibitem{Moore:1993ns}
G.~W.~Moore,
``Symmetries of the bosonic string S matrix,''
arXiv:hep-th/9310026.

%\cite{Isberg:1993av}
\bibitem{Isberg:1993av}
J.~Isberg, U.~Lindstrom, B.~Sundborg and G.~Theodoridis,
%``Classical and quantized tensionless strings,''
Nucl.Phys. B411 (1994) 122

%\cite{Ferrara:1998jm}
\bibitem{Ferrara:1998jm}
S.~Ferrara and C.~Fronsdal,
%``Gauge fields as composite boundary excitations,''
Phys.\ Lett.\ B {\bf 433} (1998) 19

%\cite{Bengtsson:1983pg}
\bibitem{Bengtsson:1983pg}
A.~K.~Bengtsson, I.~Bengtsson and L.~Brink,
%``Cubic Interaction Terms For Arbitrarily Extended Supermultiplets,''
Nucl.\ Phys.\ B {\bf 227} (1983) 41.

%\cite{Bengtsson:1983pd}
\bibitem{Bengtsson:1983pd}
A.~K.~Bengtsson, I.~Bengtsson and L.~Brink,
%``Cubic Interaction Terms For Arbitrary Spin,''
Nucl.\ Phys.\ B {\bf 227} (1983) 31.

%\cite{Vasiliev:1999ba}
\bibitem{Vasiliev:1999ba}
M.~A.~Vasiliev,
``Higher spin gauge theories: Star-product and AdS space,''\\
hep-th/9910096.

\bibitem{Savvidy:1998sb}
G.~K.~Savvidy,
%``Gonihedric string equation,''
Phys.\ Lett.\ B {\bf 438} (1998) 69


%\cite{Sezgin:2001zs}
\bibitem{Sezgin:2001zs}
E.~Sezgin and P.~Sundell,
%``Doubletons and 5D higher spin gauge theory,''
JHEP {\bf 0109} (2001) 036

%\cite{Mikhailov:2002bp}
\bibitem{Mikhailov:2002bp}
A.~Mikhailov,
%``Notes on higher spin symmetries,''
arXiv:hep-th/0201019.

%\cite{Klebanov:2002ja}
\bibitem{Klebanov:2002ja}
I.~R.~Klebanov and A.~M.~Polyakov,
%``AdS dual of the critical O(N) vector model,''
Phys.\ Lett.\ B {\bf 550} (2002) 213


%\cite{Francia:2002aa}
\bibitem{Francia:2002aa}
D.~Francia and A.~Sagnotti,
%``Free geometric equations for higher spins,''
Phys.\ Lett.\ B {\bf 543} (2002) 303


\bibitem{gon1}G.K.Savvidy and K.G.Savvidy, Mod.Phys.Lett. A8
(1993) 2963\\G.K.Savvidy, JHEP 0009 (2000) 044\\
R.V.Ambartzumian and et al, Phys.Lett. B275 (1992) 99\\
G.K.Savvidy and K.G.Savvidy, Int.J.Mod.Phys. A8 (1993) 3993\\
R.~Manvelian and G.~Savvidy, Phys.Lett.B533 (2002) 138



\bibitem{witten} E.Witten, Comm.Math.Phys.117 (1988) 353\\
E.Witten, Comm.Math.Phys.118 (1988) 411
%\cite{Labastida:1988zb}
\bibitem{Labastida:1988zb}
J.~M.~Labastida, M.~Pernici and E.~Witten,
%``Topological Gravity In Two-Dimensions,''
Nucl.Phys. B310  (1988) 611.

%\cite{Montano:1988dr}
\bibitem{Montano:1988dr}
D.~Montano and J.~Sonnenschein,
%``Topological Strings,''
Nucl.\ Phys.\ B {\bf 313} (1989) 258.


\bibitem{eguchi}T.Eguchi and S.-K.Yang, Mod.Phys.Lett. A5
(1990) 1693

\bibitem{polykov}A.M.Polyakov. Nucl.Phys.B268 (1986) 406\\
H.Kleinert. Phys.Lett. 174B (1986) 335\\
T.Curtright and P. van Nieuwenhuizen. Nucl.Phys.B294 (1987)125\\
U.Lindstr\"om,M. Ro$\breve{c}$ek and P. van Nieuwenhuizen.
Phys.Lett.B199 (1987) 219\\
U.Lindstr\"om and M. Ro$\breve{c}$ek. Phys.Lett.B201 (1988) 63

%\cite{Savvidy:dv}
\bibitem{Savvidy:dv}
G.~K.~Savvidy,
``Conformal Invariant Tensionless Strings,''
Phys.\ Lett.\ B {\bf 552} (2003) 72.


%\cite{DeVega:1992tm}
\bibitem{DeVega:1992tm}
H.~J.~De Vega and A.~Nicolaidis,
%``Strings in strong gravitational fields,''
Phys.\ Lett.\ B {\bf 295} (1992) 214.


%\cite{Clark:2003wk}
\bibitem{Clark:2003wk}
A.~Clark, A.~Karch, P.~Kovtun and D.~Yamada,
``Construction of bosonic string theory on infinitely curved Anti-de Sitter space,''
arXiv:hep-th/0304107.



\bibitem{ramond}P.Ramond, Phys.Rev.D3 (1971) 2415
\bibitem{neveu}A.Neveu and J.Schwarz, Nucl.Phys.B31 (1971) 86
\bibitem{zumino}J.L.Gervais and B.Sakita, Nucl.Phys.B34 (1971) 632\\
Y.Iwasaki and K.Kikkawa,Phys.Rev.D8 (1973) 440\\
B.Zumino, "Relativistic Strings and Supergauges"
pp.367-381 in "Renormalisation and Invariance in QFT" ed. E.Caianiello
(Plenum Press, 1974)\\
L.Brink,P. Di Vecchia and P.Howe, Phys.Lett.65B (1976) 471\\
S.Deser and B.Zumino, Phys.Lett.65B (1976) 369\\
A.M.Polyakov. Phys.Lett.103B (1981) 207; Phys.Lett.103B (1981) 211

\bibitem{Nichols:2002ux}
A.~Nichols, R.~Manvelyan and G.~K.~Savvidy,
``New strings with world-sheet supersymmetry,
hep-th/0212324.

\bibitem{Ademollo:1974fc}
M.~Ademollo, A.~D'Adda, R.~D'Auria, E.~Napolitano, P.~Di Vecchia,
F.~Gliozzi and S.~Sciuto,
%``Unified Dual Model For Interacting Open And Closed Strings,''
Nucl.Phys.  B77 (1974) 189,
Nucl.Phys.  B114 (1976) 297; Phys.Lett.62B (1976) 105\\
L.Brink and J.H.Schwarz,Nucl.Phys.B121 (1977) 285
%\cite{Ademollo:1974fc}

%\cite{D'Adda:1987rx}
\bibitem{D'Adda:1987rx}
A.~D'Adda and F.~Lizzi,
%``Space Dimensions From Supersymmetry
%For The N=2 Spinning String: A Four-Dimensional Model,''
Phys.\ Lett.\ B {\bf 191}, 85 (1987).

\bibitem{Ooguri:1990ww}
H.~Ooguri and C.~Vafa,
%``Selfduality And N=2 String Magic,''
Mod.\ Phys.\ Lett.\ A {\bf 5}, 1389 (1990).

\bibitem{Ooguri:1991fp}
H.~Ooguri and C.~Vafa,
%``Geometry of N=2 strings,''
Nucl.\ Phys.\ B {\bf 361}, 469 (1991).



\bibitem{Bershadsky:1993cx}
M.~Bershadsky, S.~Cecotti, H.~Ooguri and C.~Vafa,
%``Kodaira-Spencer theory of gravity and
%exact results for quantum string amplitudes,''
Commun.\ Math.\ Phys.\  {\bf 165}, 311 (1994)


\bibitem{Bershadsky:1993ta}
M.~Bershadsky, S.~Cecotti, H.~Ooguri and C.~Vafa,
%``Holomorphic anomalies in topological field theories,''
Nucl.\ Phys.\ B {\bf 405}, 279 (1993)

\bibitem{Ooguri:1995cp}
H.~Ooguri and C.~Vafa,
%``All loop N=2 string amplitudes,''
Nucl.\ Phys.\ B {\bf 451}, 121 (1995)


\bibitem{green}M.B.Green, J.H.Schwarz and E.Witten, Superstring Theory.
Vol.1,2 Cambridge: Cambridge University Press (1997).
\bibitem{polchinski} J.Polchinski, String Theory. Vol.1,2
Cambridge: Cambridge University Press (1998).



\bibitem{pauli}M. Fierz and W. Pauli,
Proc.Roy.Soc. A173 (1939) 211\\
W.Rarita and J.Schwinger, Phys.Rev. 60 (1941) 61


\end{thebibliography}
\end{document}